 \documentclass[final,5p,times,twocolumn,numbers]{elsarticle}

\usepackage{amssymb}
\usepackage{booktabs}

\usepackage[colorlinks=true, linkcolor=blue, citecolor=red, urlcolor=magenta]{hyperref}

\journal{Nuclear Physics B}

\begin{document}

\begin{frontmatter}



\title{Effective chemical potential and its phenomenological implications for the Hubble parameter}


\author[first]{L. L. Sales}
\ead{lazarolima@uern.br}

\author[first]{F. C. Carvalho}
\ead{fabiocabral@uern.br}

\affiliation[first]{organization={Departamento de F\'isica, Universidade do Estado do Rio Grande do Norte},
            addressline={Av. Prof. Ant\^onio Campos}, 
            city={Mossor\'o},
            postcode={59610-210}, 
            state={RN},
            country={Brazil}}

\begin{abstract}
In cosmological models, the Hubble parameter is determined by the time evolution of the scale factor, and current observations reveal a persistent tension between its values inferred from different probes, such as Cepheid variable stars and the cosmic microwave background. Within Tsallis' statistical framework, we identify two distinct definitions of fugacity associated with relativistic and non-relativistic regimes. For the non-relativistic sector, we introduce an effective chemical potential and establish its connection with the Gibbs free energy. We then explore a phenomenological correspondence between this effective chemical potential and an Unruh-like temperature associated with accelerated trajectories in an expanding cosmological background. As an application, we derive an effective expression for the Hubble parameter that includes a statistics-dependent contribution arising from the non-relativistic matter sector. This contribution suggests that non-Gaussian statistical effects, when consistently incorporated in the non-relativistic matter sector, can enhance the sensitivity of the expansion rate to underlying thermostatistical assumptions, achieving a substantial increase in sensitivity by approximately ten orders of magnitude when compared with previous relativistic constructions that investigated, at a phenomenological level, the discrepancy observed in current determinations of the Hubble constant.
\end{abstract}




\end{frontmatter}




\section{Introduction}

In the 1960s, Rindler pioneered the study of uniformly accelerated frames within the context of special relativity \cite{rindler1966kruskal}. This line of investigation led to a remarkable theoretical result known as the Unruh effect \cite{unruh1976notes}. For an accelerating observer, the lowest energy state is no longer perceived as the Minkowski vacuum; instead, it appears as a thermal bath. The temperature associated with this effect is proportional to the observer’s proper acceleration \(\alpha\) and is given by \cite{unruh1998acceleration}
\begin{equation}
	T_U = \frac{\hbar\alpha}{2\pi k_{B}c},
\end{equation}
where \(c\), \(k_B\), and \(\hbar\) denote the speed of light, the Boltzmann constant, and the reduced Planck constant, respectively. This temperature is commonly referred to as the Unruh temperature.

Despite its fundamental theoretical relevance, a direct experimental confirmation of the Unruh effect remains challenging. Consequently, several studies have explored Unruh-like phenomena in different physical settings. For instance, Rodriguez-Laguna \textit{et al.} simulated Unruh-like effects using ultracold atoms in optical lattices as quantum simulators of Dirac fields in the vicinity of an effective event horizon \cite{rodriguez2017synthetic}. It is worth emphasizing that a substantial body of literature discusses the extent to which such analog systems capture the essential features of the Unruh effect itself, including the role of space-like separation in the Unruh effect and observer-dependent horizons. In Ref.~\cite{arrechea2021inversion}, it was shown that a uniformly accelerated quantum detector in arbitrary dimensions, coupled to a field initially in its vacuum state, also exhibits Unruh-like behavior. Related effects have been investigated in the context of detectors prepared in superpositions of accelerations \cite{barbado2020unruh,foo2020unruh}. In addition to these theoretical developments, several experimental proposals aiming to probe the Unruh effect have been put forward (see, e.g., \cite{cozzella2017proposal,vriend2021unruh}). A comprehensive review of the Unruh effect and its applications can be found in Ref. \cite{crispino2008unruh}.

In recent years, significant improvements in observational precision have revealed tensions among key parameters of the standard cosmological model (\(\Lambda\)CDM). The most prominent of these is the discrepancy between independent determinations of the Hubble constant \(H_0\), currently at the level of \(4\sigma\)–\(6\sigma\) \cite{di2021realm}. Local measurements based on Type Ia supernovae calibrated with Cepheid variables favor values clustered around \(73~\mathrm{km\,s^{-1}\,Mpc^{-1}}\), whereas estimates inferred from observations of the cosmic microwave background (CMB) yield a lower value of approximately \(67.4~\mathrm{km\,s^{-1}\,Mpc^{-1}}\) \cite{aghanim2020planck}.

Recent works \cite{soares2023using,zamora2022thermodynamically} have suggested that alternative theoretical approaches may shed light on cosmological observables and their apparent discrepancies. In this context, Ref.~\cite{soares2023using} proposed a connection between the Unruh temperature and an ensemble fugacity defined through a generalized chemical equilibrium condition, leading to a statistics-dependent contribution to the Hubble parameter. Following a related strategy, the present work adopts a phenomenological approach in which accelerated, non-comoving trajectories are associated with effective thermodynamic scales. Within this framework, we determine the Hubble parameter by exploiting a correspondence between proper acceleration and statistical quantities, without invoking the detection of physical Unruh radiation or particle production. Our main new ingredient is an effective chemical potential naturally defined from the effective fugacity in the non-relativistic regime within Tsallis statistics. 

This construction provides an alternative statistics-dependent contribution to the Hubble parameter, allowing us to assess how non-equilibrium statistical effects in the non-relativistic matter sector at the level of an effective description may enhance the sensitivity of cosmological observables to underlying statistical assumptions. It is worth stressing that other connections are possible and that our choice is guided by dimensional consistency and its ability to connect acceleration-induced energy scales with the effective non-relativistic chemical potential.

This paper is organized as follows. In Section~\ref{sect2}, we discuss chemical equilibrium conditions, revisiting the concept of chemical potential in classical thermodynamics and introducing the effective chemical potential within Tsallis statistics. Our phenomenological proposal for determining the Hubble parameter through the connection between the Unruh temperature and the effective chemical potential is presented in Section~\ref{sect3}. Finally, our conclusions are summarized in Section~\ref{sect4}.

\section{Effective chemical equilibrium} \label{sect2}

\subsection{Chemical potential in classical thermodynamics}

In 1878, Josiah Willard Gibbs formulated the theory of thermodynamic equilibrium in systems that include multiple phases, such as solids, liquids, and gases. He proposed the concept of thermodynamic potentials, which are mathematical functions that define the properties of the system at equilibrium \cite{gibbs1878equilibrium}. One of these potentials is the chemical potential which, in the field of thermodynamics, refers to the energy that can either be absorbed or released due to a change in the number of particles in a chemical reaction or phase transition. This physical quantity is defined as the rate of change in the free energy of a thermodynamic system, as the number of atoms or molecules of the species are added to the system in a mixture \cite{gibbs1878equilibrium,atkins2023atkins}.

Mathematically, the chemical potential can be defined in terms of the Gibbs free energy, based on the differential relation for $dG$ \cite{pathria2016statistical,salinas2001introduction}:
\begin{equation}
	dG = -SdT + VdP + \sum_{i=1}^{n}\mu_{i}dN_{i}~.
\end{equation}
Hence, the chemical potential is given by
\begin{equation} \label{cp-definition}
	\mu_{i} = \left( \frac{\partial G}{\partial N_i}\right)_{T,P,N_{j\neq i}}~.
\end{equation}

The Gibbs free energy change of a system held at constant temperature and pressure is simply 
\begin{equation}
	dG = \sum_{i=1}^{n}\mu_{i}dN_{i}~.
\end{equation}
In a state of thermodynamic equilibrium, the Gibbs free energy is at a minimum, which means that $dG=0$. Thus, we can write
\begin{equation}
	\mu_{1}dN_{1} + \mu_{2}dN_{2} + \cdots = 0~.
\end{equation}
This allows us to establish the chemical equilibrium condition and equilibrium constant for a chemical reaction \cite{atkins2023atkins}. For instance, for $N=N_1+N_2$ with fixed $N$, we find $dN_1=-dN_2$ which implies $\mu_{1}=\mu_{2}$. This portrays the chemical balance of the system.

Understanding the role of chemical potential in a chemical reaction is crucial to knowing the direction in which a reaction happens. For the sake of illustration, consider the primordial hydrogen recombination reaction: $p+e^{-} \leftrightarrow H+\gamma$; if $\mu_{H}>\mu_{e^{-}}+\mu_{p}$, it means that the reaction is occurring preferentially from the highest energy state to the lowest one. In contrast, if $\mu_{H}<\mu_{e^{-}}+\mu_{p}$, the reaction proceeds favorably in the opposite direction.

\subsection{Effective chemical potential}

Tsallis statistics have been used successfully to explain non-equilibrium states or long-range interactions and correlations. For example, observations of the $q$-triplets by Voyager 1 in the solar wind suggest non-Gaussian effects due to non-equilibrium states \cite{burlaga2005triangle}. The long-range interactions present in several physical phenomena indicate that Tsallis' statistical mechanics achieves results further the scope of classical statistics (see, e.g., \cite{cirto2018validity,wilk2000interpretation,plastino2022tsallis,bagchi2016sensitivity,christodoulidi2016dynamics}). Likewise, recent publications have reported new observational evidence suggesting that $q$-thermostatistics accurately describes certain aspects of astrophysical self-gravitating systems (see, e.g., \cite{sanchez2022principle,almeida2021physically}).  Here, we aim to investigate non-Gaussian effects in defining chemical potential and equilibrium conditions.

In the realm of Tsallis statistics, we have shown in Ref \cite{sales2022non} that the particle number $q$-density for $q>1$ and $q<1$ reads as
\begin{equation} \label{q-density}
	n_{i}^{q} = \frac{g_{i}B_{q}}{\lambda_i^{3/2}}\left[ e_{q}^{\beta(\mu_{i}-m_{i}c^2)}\right]^{\frac{5-3q}{2}}~,
\end{equation}
where $\lambda_{i}=2\pi\hbar^{2}\beta/m_{i}$ is the thermal particle wavelength, $e_{q}^{x}=[1+(1-q)x]^{1/(1-q)}$ the $q$-exponential, and
\begin{eqnarray} 
	B_{q} = \left\{
	\begin{array}{rcl}
		\displaystyle{\frac{1}{(q-1)^{3/2}}\frac{\Gamma\left(\frac{5-3q}{2(q-1)}\right)}{\Gamma\left(\frac{1}{q-1}\right)}},& \mbox{if} & 1<q<5/3 \\ \\
		\displaystyle{\frac{1}{(1-q)^{3/2}}\frac{\Gamma\left(\frac{2-q}{1-q}\right)}{\Gamma\left(\frac{7-5q}{2(1-q)}\right)}}, & \mbox{if} & q<1~.
	\end{array}
	\right.
\end{eqnarray}
Besides, $m_i$ is the rest mass, $\mu_i$ the chemical potential and $g_i$ the degeneracy of the species $i$. Here $\beta=1/k_{B}T$, with $T$ being the bulk temperature of the system under consideration \cite{mitra2018thermodynamics}.

Taking $n_{i}^{q}$ for $\mu\neq 0$ and $\mu=0$, we have obtained the following expression \cite{sales2022non}:
\begin{eqnarray} \label{nin0q}
	\frac{n_{i}^{q}}{n_{i}^{(0),q}} = \left( e_{q}^{\theta_{i}^{q}} \right)^{\frac{5-3q}{2}} \;,
\end{eqnarray}
where
\begin{eqnarray}
	\theta_{i}^{q} = \frac{\beta\mu_{i}}{1-(1-q)\beta m_{i}c^2}\;.
\end{eqnarray}
A comparison of equation (\ref{nin0q}) with the standard fugacity definition ($z=e^{\beta\mu}$) allows us to define the effective chemical potential for non-relativistic particles as follows:
\begin{equation} \label{effective-cp}
	\mu_{i}^{q} = \frac{\mu_{i}}{1-(1-q)\beta m_{i}c^2}\;.
\end{equation}
This new generalized chemical potential depends on the particle rest energy, the $q$-parameter, and the temperature. In another way, we can write the effective chemical potential as a function of Gibbs free energy by comparing equations (\ref{cp-definition}) and (\ref{effective-cp}). It reads
\begin{equation} 
	\mu_{i}^{q} = \frac{1}{1-(1-q)\beta m_{i}c^2}\left( \frac{\partial G}{\partial N_i}\right)\;.
\end{equation}

\begin{figure}[!ht]
	\centering
	\includegraphics[width=0.8\linewidth]{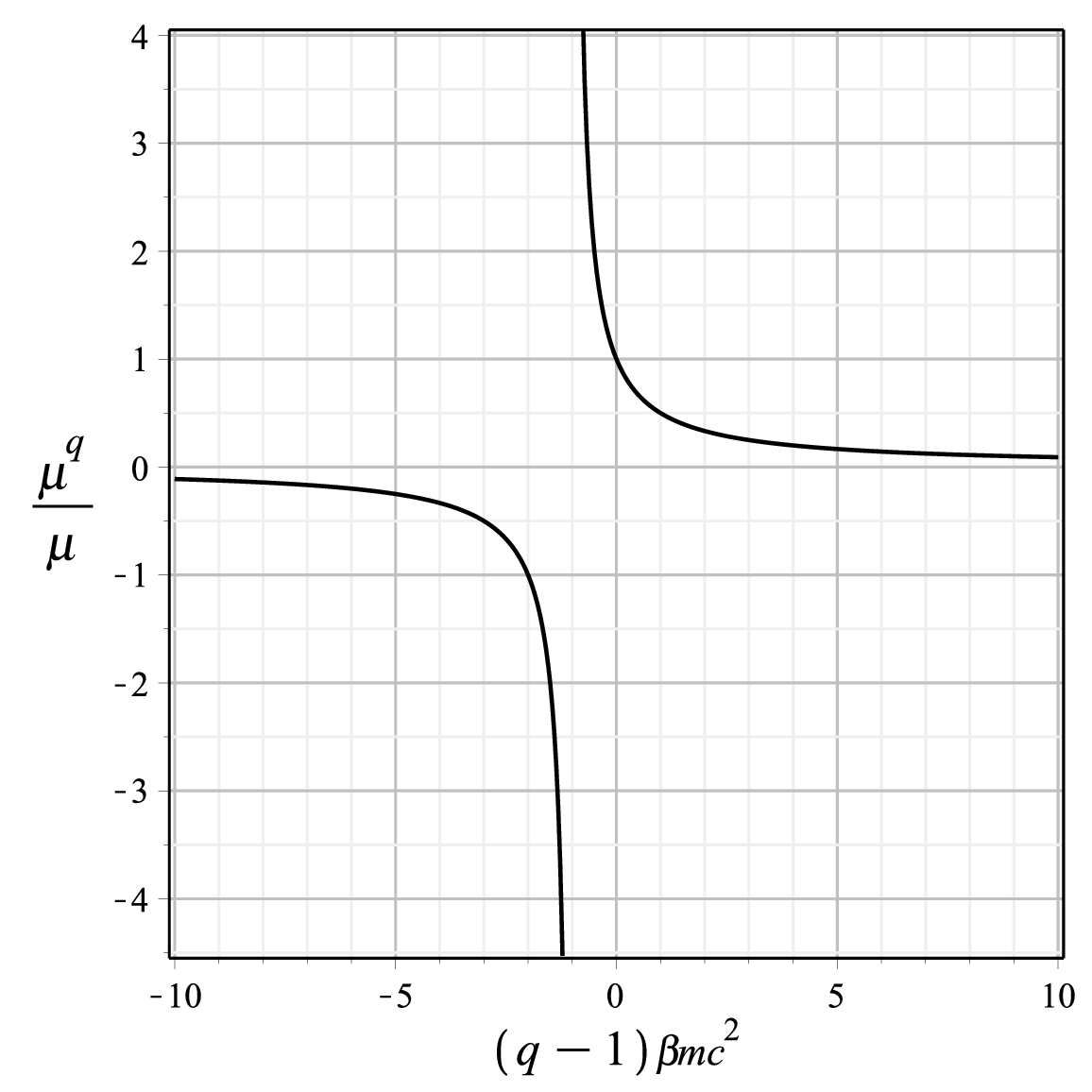}
	\caption{Comparison between the effective and standard chemical potentials for variations of $(q-1)\beta mc^2$.}
	\label{fig:q-potential}
\end{figure}

The graphical behavior of the ratio $\mu^{q}/\mu$ as a function of $(q-1)\beta mc^2$ is shown in Fig. \ref{fig:q-potential}. In this way, we can compare the effective chemical potential to the ordinary chemical potential. Note that to obtain the equality $\mu_{i}^{q}=\mu_{i}$, the following condition must be met:
\begin{equation}
	(q-1)\beta m_{i}c^2 = 0~,
\end{equation}
which leads to two solutions: $q=1$ and $\beta m_{i}c^2=0$. This latter implies the relativistic limit $k_{B}T\gg m_{i}c^2$. In other words, the definition of chemical potential remains unchanged by the statistical parameter $q$ in the relativistic regime. Regardless, as we will see later, the chemical equilibrium condition in this energy range is affected by the $q$-parameter. On the other hand, when $\mu_{i}^{q}=-\mu_{i}$, the $q$-parameter is given by
\begin{equation}
	q = 1 - \frac{2}{\beta m_{i}c^2}~.	
\end{equation}	

As can be seen, the graph in Fig \ref{fig:q-potential} presents a singularity at $(q-1)\beta mc^2=-1$. This leads to the following constraint for the $q$-parameter:
\begin{equation} \label{rpq}
	q \neq 1 - \frac{1}{\beta mc^2}~.
\end{equation}
Furthermore, the graph has the shape of a hyperbola, so that 
\begin{equation}
	\frac{\mu^{q}}{\mu} < 0~~~~\textrm{if}~~~~q < 1 - \frac{1}{\beta mc^2}~,
\end{equation}
and
\begin{equation}
	\frac{\mu^{q}}{\mu} > 0~~~~\textrm{if}~~~~q > 1 - \frac{1}{\beta mc^2}~.
\end{equation}

In reference \cite{sales2022non}, we have established the non-Gaussian Saha ionization equation for the primordial hydrogen recombination reaction, i.e., $p+e^{-} \leftrightarrow H+\gamma$. There, we have shown that the chemical equilibrium condition under which Saha's equation is built yields
\begin{equation} \label{theta_eq}
	\theta_{H}^{q} = \theta_{e}^{q} + \theta_{p}^{q} + (1-q)\theta_{e}^{q}\theta_{p}^{q}\;.
\end{equation}
However, with this paper's effective chemical potential definition, we can express the above equation using this new potential as
\begin{equation} \label{q-potencial}
	\mu_{H}^{q} = \mu_{e}^{q} + \mu_{p}^{q} + (1-q)\beta \mu_{e}^{q}\mu_{p}^{q}\;.
\end{equation}
Considering the approximation $m_{H}\approx m_{p}$, equation (\ref{q-potencial}) provides the effective chemical equilibrium condition for non-relativistic particles in terms of the ordinary chemical potential \cite {sales2022non}, given by
\begin{equation} \label{ceqg}
	\mu_{H} = \sigma_{q}\mu_{e} + \mu_{p} + (1-q)\sigma_{q}\beta\mu_{e}\mu_{p}\;,
\end{equation}
where
\begin{equation} 
	\sigma_{q} = \frac{1-(1-q)\beta m_{p}c^2}{1-(1-q)\beta m_{e}c^2}\;.
\end{equation} 

An important aspect to highlight is the non-Gaussian effect in the equilibrium condition, which is influenced by both the $q$-parameter and the temperature. The contribution of the species' rest energy is also observed, as evidenced by the factor $\sigma_q$. The constraints for $q$, according to three conditions for the factor $\sigma_q$, are presented in Table \ref{q-restricao}. Note that if $\sigma_q=0$, then $q=1-1/\beta m_p c^2$, which violates the restriction on the effective chemical potential given in equation (\ref{rpq}). Hence, the factor $\sigma_q$ cannot be equal to zero, while for $\sigma_q>0$ and $\sigma_q<0$, the constraints for the $q$-parameter depend on the rest energies of electrons and protons as well as the temperature.

\begin{table}[!ht] 
	\begin{center}
		\caption{Constraints on the $q$-parameter for the positive, negative, and null $\sigma_{q}$ factor.}
		\label{q-restricao}
		\begin{tabular}{cc} 
			\toprule
			Condition & Constraints on $q$ \\ \midrule
			$\sigma_{q}>0$ &  $\displaystyle{1 - \frac{1}{\beta m_e c^2}>q>1 - \frac{1}{\beta m_p c^2}}$ \\ 
			$\sigma_{q}<0$ &  $\displaystyle{1 - \frac{1}{\beta m_e c^2}<q<1 - \frac{1}{\beta m_p c^2}}$ \\ 
			$\sigma_{q}=0$ & $\displaystyle{q = 1 - \frac{1}{\beta m_p c^2}}$ \\ \bottomrule
		\end{tabular}
	\end{center} 
\end{table}

Now, let us assess how the $q$-parameter affects the direction in which the recombination reaction occurs. For example, for $\beta\mu_{p}>0$, if
\begin{equation} \label{rq1}
	q < 1 + \frac{1}{\beta\mu_{p}}\left( 1-\frac{1}{\sigma_{q}}\right)~~~~(\sigma_{q}>0)~,
\end{equation}
and
\begin{equation} \label{rq2}
	q > 1 + \frac{1}{\beta\mu_{p}}\left( 1-\frac{1}{\sigma_{q}}\right)~~~~(\sigma_{q}<0)~,
\end{equation}
we shall have $\mu_{H}>\mu_{p}+\mu_{e}$, i.e., the reaction will proceed from the highest energy state to the lowest energy state and will happen in the opposite direction, $\mu_{H}<\mu_{p}+\mu_{e}$, whenever
\begin{equation} \label{rq3}
	q > 1 + \frac{1}{\beta\mu_{p}}\left( 1-\frac{1}{\sigma_{q}}\right)~~~~(\sigma_{q}>0)~,
\end{equation}
and
\begin{equation} \label{rq4}
	q < 1 + \frac{1}{\beta\mu_{p}}\left( 1-\frac{1}{\sigma_{q}}\right)~~~~(\sigma_{q}<0)~.
\end{equation}

If we consider the relativistic limit ($k_{B}T\gg m_{i}c^2$), we have $\sigma_{q}\approx 1$, and then equation (\ref{ceqg}) leads to a chemical equilibrium condition for relativistic particles \cite{sales2022non}, namely, 
\begin{equation} \label{ceqgr}
	\mu_{H} = \mu_{p} + \mu_{e} + (1-q)\beta\mu_{e}\mu_{p}\;.
\end{equation}
Another pathway to achieve the same chemical equilibrium state as in equation (\ref{ceqgr}) is by solving for the $q$-number density in the relativistic regime. It follows 
\begin{equation} \label{dnpr}
	n_{i}^{q} = \frac{4\pi g_{i}}{(2\pi\hbar)^{3}}e_{q}^{\beta\mu_{i}}\int_{0}^{b} dpp^{2}\left[1+(q-1)\alpha_i p\right]^{\frac{1}{1-q}}~,
\end{equation}
where
\begin{eqnarray} \label{b}
	b = \left\{
	\begin{array}{rl}
		\infty, & q>1 \\
		\left[(1-q)\alpha_{i}\right] ^{-1}, & q<1~.
	\end{array}
	\right.
\end{eqnarray}
The parameter $\alpha_i$ is defined as
\begin{equation} 
	\alpha_i = \frac{\beta c}{1+(1-q)\beta\mu_{i}}~.
\end{equation}
Evaluating the integral in equation (\ref{dnpr}), we obtain 
\begin{equation} 
	n_{i}^{q} = \frac{g_i}{\pi^2}\left(\frac{k_{B}T}{\hbar c}\right)^3 G_{q}\left[ e_{q}^{\beta\mu_i} \right]^{4-3q}~,
\end{equation}
where
\begin{eqnarray} 
	G_{q} = \left\{
	\begin{array}{rcl}
		\displaystyle{\frac{1}{(q-1)^{3}}\frac{\Gamma\left(\frac{4-3q}{q-1}\right)}{\Gamma\left(\frac{1}{q-1}\right)}},& \mbox{if} & 1<q<4/3 \\ \\
		\displaystyle{\frac{1}{(1-q)^{3}}\frac{\Gamma\left(\frac{2-q}{1-q}\right)}{\Gamma\left(\frac{5-4q}{1-q}\right)}}, & \mbox{if} & q<1~.
	\end{array}
	\right.
\end{eqnarray}

Therefore, a ratio analogous to equation (\ref{nin0q}) for relativistic particles can be set as follows:
\begin{equation} \label{nin0q2}
	\frac{n_{i}^{q}}{n_{i}^{(0),q}} = \left( e_{q}^{\beta\mu_{i}} \right)^{4-3q} \;.
\end{equation}
Performing the manipulations with the $q$-exponential and $q$-logarithm functions, together with equation (6) from reference \cite{sales2022non}, it is straightforward to show that equation (\ref{ceqgr}) is obtained.

Notice that, in equation (\ref{ceqgr}), we shall have the inequality $\mu_{H}>\mu_{p}+\mu_{e}$ for $q<1$ and $\mu_{H}< \mu_{p}+\mu_{e}$ if $q>1$. These conditions for the $q$-parameter are also easily attained when we insert $\sigma_{q}\approx 1$ into equations (\ref{rq1}) and (\ref{rq3}). It is worth stressing that these comparisons are made regarding standard chemical equilibrium. In summary, our findings suggest that the $q$-parameter plays a role in controlling the direction in which the reaction occurs in both energy regimes. Moreover, since the chemical potential is a form of potential energy, the cross terms $\mu_{e}\mu_{p}$ in equations (\ref{ceqg}) and (\ref{ceqgr}) can be related to energy dissipation. 

It is convenient to define two forms of fugacity in $q$-statistics once we have obtained two chemical equilibrium conditions. In the case of a non-relativistic gas, the effective fugacity can be defined as follows:
\begin{equation} \label{qfugacityNR}
	z_{i}^{q} = \left( e_{q}^{\beta\mu_{i}^{q}} \right)^{\frac{5-3q}{2}}~~~~(m_{i}c^2\gg k_{B}T)~.
\end{equation}
In this case, the definition above was based on equation (\ref{nin0q}). Likewise, based on equation (\ref{nin0q2}), the effective fugacity for a relativistic gas can be defined as
\begin{equation} \label{nin0qr}
	z_{i}^{q} = \left( e_{q}^{\beta\mu_{i}} \right)^{4-3q}~~~~(k_{B}T\gg m_{i}c^2).
\end{equation}
Both definitions above converge to classical fugacity when $q\rightarrow 1$. Classical fugacity is a measure of how much a system deviates from the behavior of an ideal gas. Effective fugacity has two definitions that can be helpful when studying systems that are out of thermodynamic equilibrium and have long-range interactions or strong statistical correlations \cite{tsallis2023introduction}.

\section{Applications: Effective Hubble parameter} \label{sect3}

\subsection{Preliminary}

In this section we investigate phenomenological applications in a cosmological context. Previous studies have explored possible connections between the Unruh temperature and thermostatistical quantities as alternative pathways to express the Hubble parameter. In particular, Ref.~\cite{soares2023using} introduced the so-called chemical $q$-potential, defined as
\begin{equation}
	\mu_{q} = \mu_{p} + \mu_{e} + (1-q)\beta\mu_{e}\mu_{p} - \mu_{H}\, ,
\end{equation}
which can be interpreted as a generalized chemical equilibrium condition for relativistic particles, rather than as a chemical potential in the strict thermodynamical sense. Based on this relation, a phenomenological link was established between the Unruh temperature and the Hubble parameter, suggesting that statistical effects may play a role in discussions of the Hubble tension.

Within this framework, the authors proposed an alternative way to assess the discrepancy between two independent determinations of the Hubble constant, namely those inferred from Cepheid variable stars and from the CMB. Their analysis led to the estimate \cite{soares2023using}
\begin{equation} \label{eq_new}
	\Delta[(1-q)\bar{\mu}]
	= [(1-q)\bar{\mu}]_{\rm Cepheids}
	- [(1-q)\bar{\mu}]_{\rm CMB}
	\sim 10^{-52},
\end{equation}
where $\bar{\mu}$ denotes the fraction of the reduced chemical potential. The extremely small magnitude of this quantity has been interpreted as indicating that imperceptible differences in the underlying thermostatistics could be associated with the observed discrepancy in the inferred values of $H_0$.

In the present work, we follow a related strategy but introduce a distinct and physically motivated element: an effective chemical potential naturally defined in the non-relativistic regime within Tsallis statistics. This construction leads to a modified statistics-dependent contribution to the Hubble parameter, which differs from that obtained in Ref.~\cite{soares2023using}. As will be shown in the next subsection, this alternative formulation does not resolve the Hubble tension, but it significantly enhances the sensitivity of the expansion rate to non-Gaussian statistical effects, thereby highlighting their potential relevance in the phenomenological analysis of the Hubble parameter.

\subsection{Phenomenological formulation: effective non-relativistic contribution}

In this subsection, we present an alternative formulation to connect the effective chemical potential introduced in Sec.~2 with the cosmological expansion rate. Our approach is phenomenological and relies on a thermodynamic correspondence between accelerated trajectories and effective temperature scales.

\begin{figure}[ht]
	\centering
	\includegraphics[width=1.0\linewidth]{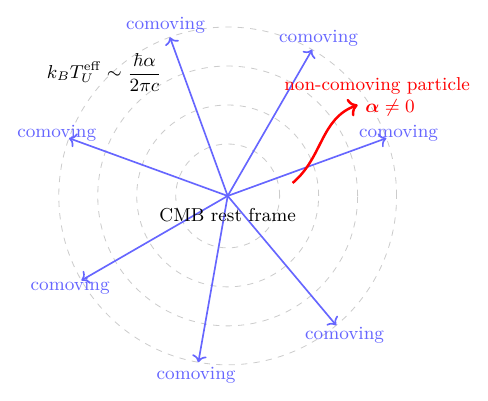}
	\caption{Schematic phenomenological illustration of non-comoving particles embedded in an expanding FLRW background. No physical Unruh radiation or particle production is implied.}
	\label{fig:flrw_unruh}
\end{figure}

We consider non-comoving particles embedded in an expanding FLRW background. Although fundamental observers comoving with the cosmic fluid follow geodesic trajectories, particles that are not exactly comoving experience a nonvanishing proper acceleration with respect to the cosmological rest frame defined by the CMB. In this context, it is natural to associate an effective Unruh-like temperature to such accelerated trajectories, not as a source of physical radiation, but as a local thermodynamic mapping between acceleration and energy scales. A schematic illustration of this configuration is shown in Fig.~\ref{fig:flrw_unruh}.

The use of a non-equilibrium statistical framework is motivated by the fact that non-comoving particles evolve in a setting dominated by long-range gravitational interactions and by the non-stationary cosmic expansion. While comoving observers can absorb the large-scale dynamics into the background geometry and admit a homogeneous equilibrium description, non-comoving observers experience a proper acceleration relative to the CMB frame, which induces an effective energy flux and generically breaks local thermodynamic equilibrium. In this context, Tsallis statistics could provide a natural effective description of the matter sector, with the entropic parameter \(q\) quantifying deviations from equilibrium associated with the observer’s trajectory. The proper acceleration then sets a local energy scale, which is phenomenologically mapped onto the effective chemical potential through the correspondence introduced below.

Assuming that the chemical potential characterizing the non-relativistic ensemble associated with such accelerated trajectories is effectively described by the non-relativistic effective chemical potential defined in Eq. (\ref{effective-cp}), we postulate the following phenomenological correspondence between an Unruh-like temperature scale and the effective chemical potential:
\begin{equation}
	\label{unruh-potential}
	\left|(1-q)\bar{\mu}_q\right|
	= \frac{k_B T_U}{m_i c^2}
	= \frac{\hbar \alpha}{2\pi m_i c^3},
	\qquad (T_U \geq 0),
\end{equation}
where $\alpha$ denotes the proper acceleration of the particle trajectory.
Here,
\begin{equation}
	\bar{\mu}_q
	= \frac{\mu_i^{q}}{\mu_i^{q}-\mu_i}
	= \frac{1}{(1-q)}\frac{k_B T}{m_i c^2}
\end{equation}
represents the fraction of the reduced effective chemical potential. In this work, the $q$-parameter should be interpreted as an effective parameter dependent on the physical environment or on the class of non-comoving trajectories considered.

At this stage, it is crucial to distinguish between the different notions of temperature involved in this construction. The temperature entering the effective chemical potential is the statistical temperature of the ensemble, defined through the effective fugacity within Tsallis' framework. This quantity should not be identified with the Unruh temperature, which originates from the kinematic association between proper acceleration and energy scales in quantum field theory. In the present context, the Unruh-like temperature is not interpreted as a physical thermodynamic temperature associated with particle production or radiation. Rather, it is employed as a local energetic scale induced by the acceleration of non-comoving particles with respect to the cosmological rest frame. The identification between the statistical temperature of the ensemble and the Unruh-like temperature is therefore assumed only at a phenomenological level, allowing one to relate acceleration effects to the effective chemical potential without invoking thermal equilibrium.

In the non-relativistic and sub-horizon regime, the relative velocity between a non-comoving particle and a comoving observer can be approximated kinematically by the local Hubble flow. To relate the acceleration to the cosmological expansion, we consider the kinematics implied by Hubble's law, $v = Hr$. Taking its time derivative, one finds
\begin{equation}
	\label{dHl}
	\frac{\alpha}{r} = \dot{H} + H^2,
\end{equation}
where $\alpha = \dot{v}$. This relation can be interpreted as the relative acceleration between nearby trajectories in an expanding homogeneous and isotropic universe and is consistent with the Raychaudhuri equation for a congruence of worldlines.

Using the correspondence introduced in Eq.~(\ref{unruh-potential}), the ratio
$\alpha/r$ may be written as
\begin{equation}
	\label{alpha-r-ratio2}
	\frac{\alpha}{r}
	= \frac{\ddot{a}}{a}
	= \frac{2\pi m_i c^3}{\hbar r_0}\frac{(1-q)\bar{\mu}_q}{a},
\end{equation}
where $r(t)=a(t)r_0$ with $r_0$ being a characteristic comoving length scale.
Combining Eqs.~(\ref{dHl}) and (\ref{alpha-r-ratio2}), we obtain
\begin{equation}
	\label{H2}
	H^2
	= \frac{2\pi m_i c^3}{\hbar r_0}\frac{(1-q)\bar{\mu}_q}{a}
	- \dot{H}.
\end{equation}

Adopting a spatially flat cosmological model ($k=0$) and using the Friedmann equations, the time derivative of the Hubble parameter can be written as
\begin{equation}
	\dot{H}
	= \frac{\ddot{a}}{a}
	- \left(\frac{\dot{a}}{a}\right)^2
	= -4\pi G\left(\rho + \frac{p}{c^2}\right),
\end{equation}
which leads to
\begin{equation}
	\label{Hubble3}
	H^2
	= 4\pi G\left(\rho + \frac{p}{c^2}\right)
	+ \frac{2\pi m_i c^3}{\hbar r_0}\frac{(1-q)\bar{\mu}_q}{a}.
\end{equation}

The second term in Eq.~(\ref{Hubble3}) represents a statistics-dependent contribution associated with the non-relativistic effective chemical potential.

It is important to stress that this term does not constitute a new gravitational source, it encodes non-equilibrium statistical corrections to the energy balance of the non-relativistic matter sector. The spacetime geometry remains fully relativistic, while the effective chemical potential applies exclusively to the matter content. In the current universe, the energy density is dominated by dark energy and non-relativistic matter, whereas the relativistic contribution from radiation and neutrinos is negligible \cite{aghanim2020planck}. This justifies the use of a non-relativistic effective chemical potential in assessing thermostatistical effects on the current expansion rate. 

The accelerated trajectories considered here may be associated with test particles, such as electrons or protons, that are not exactly comoving with the cosmological fluid. Such non-comoving behavior can arise from peculiar velocities induced by local gravitational inhomogeneities or, in the case of charged particles, from interactions with cosmic electromagnetic fields. For numerical estimates, we model the non-comoving observer as a proton, whose rest mass sets the relevant non-relativistic energy scale in the effective chemical potential. Numerical estimates should be regarded as a baryonic benchmark, not yet as a full model for dark matter or the total matter sector. Thus, equation (\ref{Hubble3}) then becomes
\begin{equation}
	\label{Hsquare3}
	H^2
	\approx 10^{-9}\rho
	+ 10^{-26}p
	+ 10^{6}\frac{(1-q)\bar{\mu}_q}{a}.
\end{equation}
Here, we adopted the following numerical values: $\hbar = 1.055\times 10^{-34}~\mathrm{J\,s}$, $G = 6.674\times 10^{-11}~\mathrm{m^3\,kg^{-1}\,s^{-2}}$, $c = 2.998\times 10^{8}~\mathrm{m\,s^{-1}}$, $r_0 = 4.4\times 10^{26}~\mathrm{m}$, and $m_p = 1.673\times 10^{-27}~\mathrm{kg}$. It is worth noting that if $r_0$ varies by an order of magnitude, the coefficient varies inversely in the same proportion. Furthermore, if we replace a proton with an electron, the coefficient drops to $m_e/m_p\sim1/1836$. Therefore, these values were used only as a reference scale.

Notably, the coefficient of the statistics-dependent term is significantly larger than those multiplying the energy density and pressure in Eq.~(\ref{Hsquare3}). Among the standard contributions, the pressure term is the least relevant, followed by the energy density. Under these considerations, we can write
\begin{equation}
	\label{Hsquare4}
	\Delta(H^2)
	\approx 10^{6}\,\Delta[(1-q)\bar{\mu}_q],
\end{equation}
where $\Delta$ denotes the difference between quantities inferred from the two observational probes. In addition, the scale factor measured today is set to $a=a_0=1$. Adopting $H_0 = 73.04~\mathrm{km\,s^{-1}\,Mpc^{-1}}$ for Cepheid-based measurements \cite{reid2019improved} and $H_0 = 67.4~\mathrm{km\,s^{-1}\,Mpc^{-1}}$ for CMB observations \cite{aghanim2020planck}, we find $\Delta(H^2)\sim 10^{-36}~\mathrm{s^{-2}}$, which yields
\begin{equation}
	\Delta[(1-q)\bar{\mu}_q]
	= [(1-q)\bar{\mu}_q]_{\rm Cepheids}
	- [(1-q)\bar{\mu}_q]_{\rm CMB}
	\sim 10^{-42}.
\end{equation}
Consequently, the ratio between the present estimate and that obtained in Ref. \cite{soares2023using}, Eq.~(\ref{eq_new}), reads
\begin{equation}
	\frac{\Delta[(1-q)\bar{\mu}_q]}{\Delta[(1-q)\bar{\mu}]}
	\sim 10^{10}.
\end{equation}

Using the Cepheid and CMB values for $H_0$, the corresponding fractional shift is approximately $8.4\%$. We emphasize, however, that our framework does not dynamically predict such a shift. In this regard, within the present effective parameterization, it provides an estimate of the quantity $\Delta[(1-q)\bar{\mu}_q]$ that would be required to reproduce a discrepancy of this order. Hence, the relevance of the result is phenomenological: although the inferred quantity remains extremely small in absolute terms, its magnitude is enhanced by about ten orders relative to the previous relativistic construction of Ref. \cite{soares2023using}.

Although the present analysis is restricted to the homogeneous background level, the effective correction discussed here could, in principle, have consequences for late-time cosmological observables. In particular, any modification of the expansion history may affect background probes such as direct measurements of $H(z)$, luminosity distances from type Ia supernovae, angular-diameter distances, and BAO-related quantities. At the current stage, however, these should be regarded only as prospective observational signatures of the framework, rather than as established predictions. A proper assessment of structure growth, CMB observables, or redshift-space distortions would require extending the model to cosmological perturbations, which lies beyond the scope of the present work.

\section{Conclusion} \label{sect4}

In this study, we have demonstrated that two distinct definitions of fugacity naturally emerge within a non-Gaussian statistical framework, depending on the energy regime of the system. In particular, we introduced an effective chemical potential applicable to non-relativistic particles and established its connection with the Gibbs free energy. We have also shown that, in the relativistic limit, an effective chemical potential cannot be consistently defined; instead, the appropriate quantities are the effective fugacity and the generalized chemical equilibrium condition. In both regimes, we derived constraints on the entropic parameter $q$ and examined its influence on the direction of chemical reactions, highlighting the role of non-Gaussian effects in non-equilibrium processes.

As an application, we explored a phenomenological correspondence between the non-relativistic effective chemical potential and an Unruh-like temperature associated with accelerated trajectories in an expanding universe. Within this framework, we derived an effective expression for the Hubble parameter and showed that the statistics-dependent contribution differs from that obtained in Ref.~\cite{soares2023using} by a factor of approximately $10^{10}$. This enhancement originates from the explicit inclusion of the non-relativistic rest-mass energy in the effective chemical potential.

Importantly, this result does not provide a dynamical solution to the Hubble tension, nor does it eliminate the discrepancy between different determinations of the Hubble constant. Rather, it indicates that non-Gaussian statistical effects, when consistently incorporated in the non-relativistic matter sector, can significantly amplify the sensitivity of the expansion rate to underlying thermostatistical assumptions. In this sense, differences in the statistical properties associated with distinct observational probes may contribute, at a phenomenological level, to the observed tension in the current values of $H_0$.

\section*{Acknowledgements}
FCC was supported by CNPq/FAPERN/PRONEM.




\bibliographystyle{elsarticle-num} 
\bibliography{example}






\end{document}